Cover Sheet

2018 ASEE Annual Conference & Exposition
Paper ID #22803
Session: Computing Technology Applications-I

Link:
https://www.asee.org/public/conferences/106/papers/22803/view

Authors:
Dr. Raja S. Kushalnagar, Gallaudet University
Mr. Gary W. Behm, Rochester Institute of Technology
Kevin T. Wolfe
Mr. Peter Yeung
Miss Becca Dingman
Mr. Shareef Sayel Ali, Center on Access Technology
Mr. Abraham Glasser, Rochester Institute of Technology
Ms. Claire Elizabeth Ryan

# RTTD-ID: Tracked captions with multiple speakers for deaf students

**Abstract**
Students who are deaf and hard of hearing cannot hear in class and do not have full access to spoken information. They can use accommodations such as captions that display speech as text. However, compared with their hearing peers, the caption accommodations do not provide equal access, because they are focused on reading captions on their tablet and cannot see who is talking. This viewing isolation contributes to student frustration and risk of doing poorly or withdrawing from introductory engineering courses with lab components.  It also contributes to their lack of inclusion and sense of belonging.

We report on the evaluation of a Real-Time Text Display with Speaker-Identification, which displays the location of a speaker in a group (RTTD-ID). RTTD-ID aims to reduce frustration in identifying and following an active speaker when there are multiple speakers, e.g., in a lab. It has three different display schemes to identify the location of the active speaker, which helps deaf students in viewing both the speaker's words and the speaker's expression and actions.

We evaluated three RTTD speaker identification methods: 1) traditional: captions stay in one place and viewers search for the speaker, 2) pointer: captions stay in one place, and a pointer to the speaker is displayed, and 3) pop-up: captions "pop-up" next to the speaker. We gathered both quantitative and qualitative information through evaluations with deaf and hard of hearing users. The users preferred the pointer identification method over the traditional and pop-up methods.

**Introduction**
In engineering classes and labs, deaf participants often access spoken information through captioners, who type the speech as text. People who hear can listen and look at the active speaker at the same time, while people who do not hear are focused on reading captions on their tablet and cannot see who is talking.

Reading captions when there are multiple speakers can be very challenging for deaf students, because speakers can rapidly take turns in any order. The deaf student is usually focused on reading the captions and cannot anticipate who will speak next [1]. As a result, the deaf student usually looks back and forth between captions and the speakers, and they become tired or distracted [2]. Additionally, deaf participants can feel left out of the conversation and as though they do not grasp the material being presented. This disconnect between hearing speakers and deaf participants can cause misunderstanding and miscommunication, and isolate deaf people from the class [3].

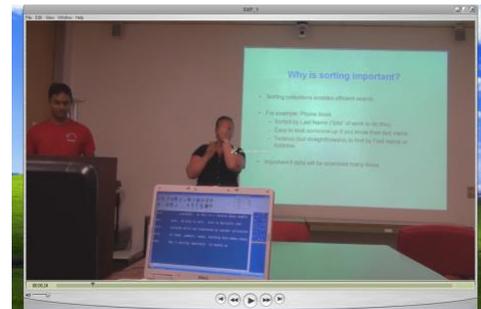
*Figure 1: Multiple information sources: multiple speakers, slides, and captions*

Deaf and hard of hearing people are underrepresented in college, especially STEM and Engineering courses. For example, even though 17% of all adults in the United States report some difficulty hearing, less than 5% of all adults employed in STEM are deaf [4]. The National Science Foundation [5] and National Academy of Engineering [6], [7] have started to encourage research initiatives in inclusion and diversification in undergraduate classes, especially in Science, Technology, Engineering, and Mathematics (STEM) fields. The NSF and NAE have made these research initiatives a priority because studies have shown that teams that are diverse are more effective at problem solving [8], making decisions [9], and ultimately have a greater impact on the quality of science produced [10]. As William Wulf, former President of the National Academy of Engineering, eloquently stated: "… creativity does not spring from nothing; it is grounded in our life experiences. Lacking diversity on an engineering team, we limit the set of solutions that will be considered and we may not find the best, the elegant solution." [6].

**Related works**

RTTD-ID builds on Real-Time Text Display (RTTD) developed by Kushalnagar, et al [11]. for classroom use. RTTD is a caption display method which tracks a single speaker moving across a classroom and projects captions transcribed by a C-Print captioner or Automatic Speech Recognition, above them, allowing deaf viewers to more closely follow what a speaker is saying. The system is designed to be portable, easy to set-up, and low-cost. It uses a Microsoft Kinect 2 to track the position of the speaker. The captions are displayed via a projector as shown in Figure 2.

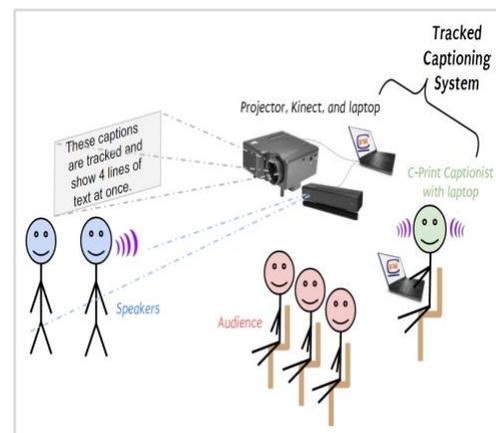

*Figure 2: RTTD with multiple speakers*

Kushalnagar et al., [11], [12] found RTTD to be an effective captioning method in the classroom setting, improving students' ability to follow along with a lecture and to understand lecture content over traditional captioning. However, the system does not indicate who is speaking.

While the traditional method of inserting the speaker's name in captions for speaker identification has been shown to be useful for viewers who are hard of hearing [13], [14], however these studies have also found that viewers who were deaf did not view it as useful.

One reason why deaf viewers may not view traditional captions as useful is that the display method does not show the location of the speaker. We address this issue by implementing two new tracking and caption display methods – moving the captions next to the speaker, or pointing to the speaker.

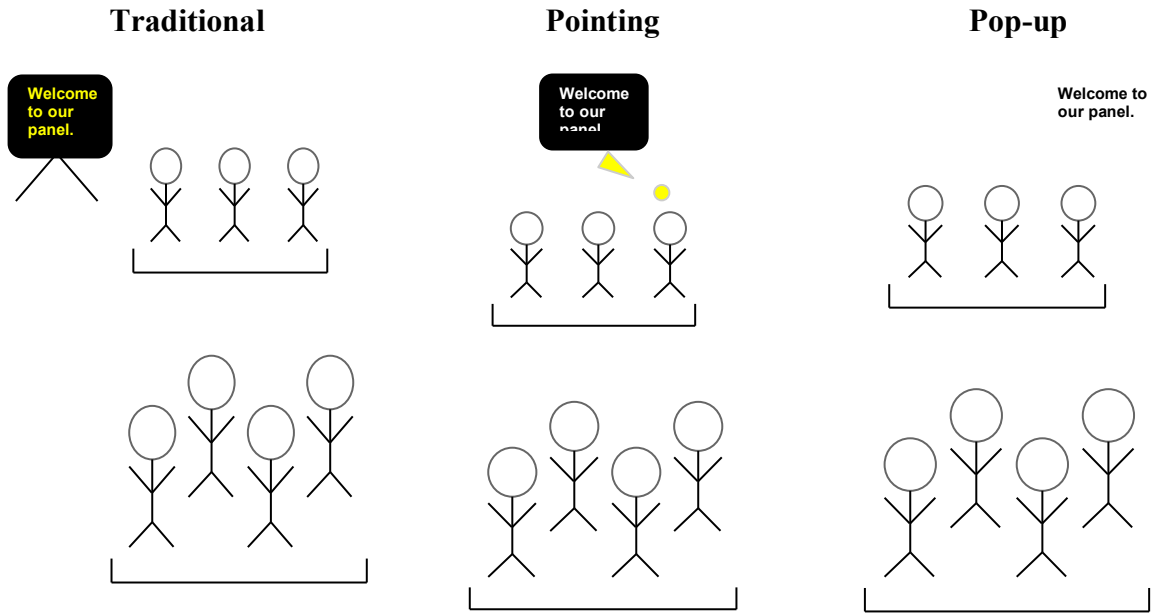

*Figure 3: Example of the different captioning styles*

**Methodology**

The RTTD-ID consists of a projector, Kinect 2, and a laptop. We developed a custom program (RTTD-ID) that collects individual locations of people in front of a Kinect 2 through its motion and audio sensors. The custom program used the tracking information to display the captions using one of the three display formats: traditional, pointing and pop-up.

The traditional captioning method displays captions in a fixed location, and the speaker's name is inserted whenever a new person speaks up. The pointing and pop-up methods either add a pointer to speaker, or move the captions next to the speaker. For the pointing and pop-up methods, the program detects which speaker raised a hand and gives that person the floor. A person raises any hand and lowers it to obtain control of the captions until another person raises a hand. This form of control based on hand raising takes advantage of social dynamics - when someone motions with a hand, others know that person would like to speak or to add something to the conversation. It is a method which reflects physical-world experiences. These methods provide a more obvious indicator of who is speaking as shown in Figure 3.

**Evaluation**

For the study, we recruited 15 participants through flyers and targeted emails on campus. Of the 15 participants, 9 participants identified as deaf, and 6 participants identified as hard-of-hearing. There were 7 participants who identified as women, and 8 participants who identified as men. The participants were at least 18 years old. Participants were compensated $20 for participating in our study.

For our study, we created three videos with the three different captioning methods. All videos had three speakers. We wanted to keep the captions same for all panels and present, and we felt that recording the three different panels ahead of time would be easiest. We used our projector to project the pre-recorded video along with the different captioning styles.

The experiment was set up with two desks and chairs for the participants as shown in Figure 4. The evaluation had three parts: a face-to-face introduction and orientation, a survey evaluation, and a feedback survey in which they were given an opportunity to ask and get answers to any questions they had about their participation. The surveys were administered immediately after they viewed the captions through the specified caption display method. The participants were assigned identification numbers to maintain confidentiality.

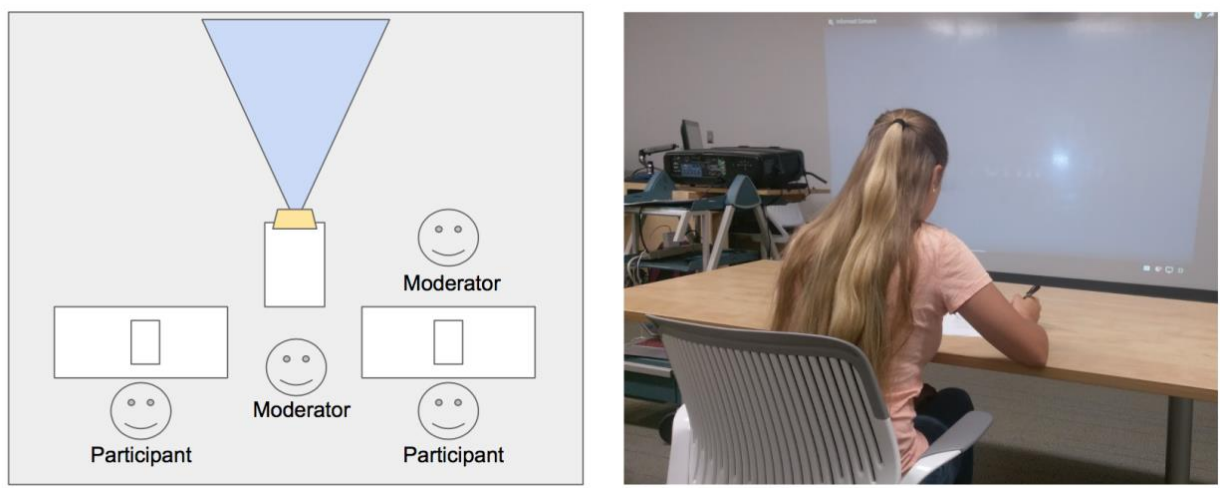

*Figure 4: A typical layout of the evaluation.*

All participants viewed three presentations in a randomized order. We randomized the order of presentations viewed by the participants to counterbalance them. After each presentation, the moderators stepped in and set up the system to display the next presentation. The first set of 5 participants viewed presentation A, "Plastic Bag Ban in Bali" first, then B, "Black Lives Matter Founders", and last, C, "Jane Fonda and Lily Tomlin: A hilarious celebration of lifelong female friendship". The next set of 5 participants viewed the presentations in a different order such as B, C, A. The last set of participants viewed the presentations in C, A, B order.

To collect audience feedback on the captioning methods for both talks, participants were asked to complete online surveys relating to the ease of following the tracked captions, the helpfulness of the tracked captions, and their engagement due to the tracked captions, as shown in Figures 5 through 10. Additionally, the individuals were asked about their perceptions on single and multi-speaker presentations through an open-ended response.

**Results**

For the open-ended question about captioned multi-speaker presentations versus single speaker presentations, all said that multi-speaker presentations were more confusing than single-speaker presentations, because they could not tell who was speaking and when they started speaking. Out of all participants, six said that they never looked at the speaker since it was too hard and tiring for them. Participant responses to the survey preference questions is shown below in Figures 5 through 9 (speaker identification, caption understanding, difficulty in following captions, fatigue in following captions, and comfortableness in following captions). Finally, Figure 10 summarizes the participants' preferences by caption method.

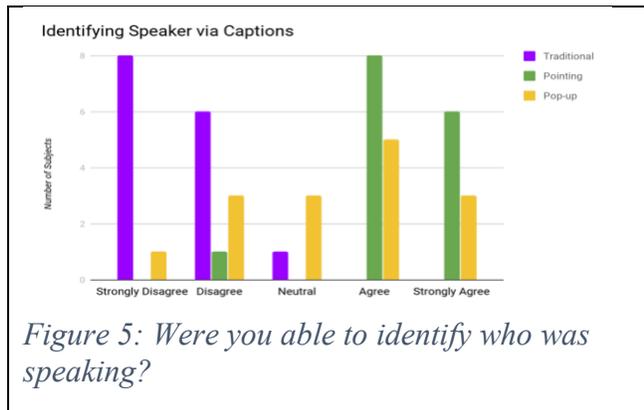

*Figure 5: Were you able to identify who was speaking?*

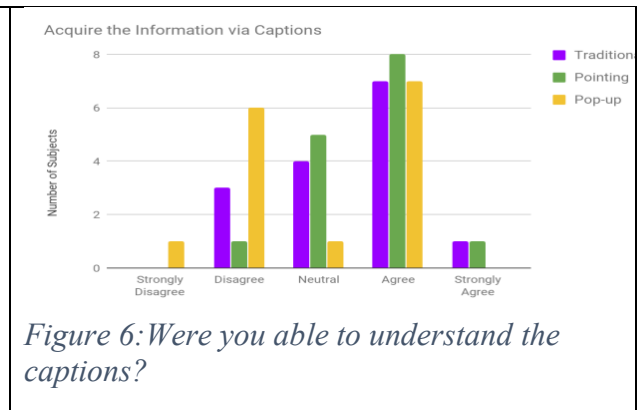

*Figure 6: Were you able to understand the captions?*

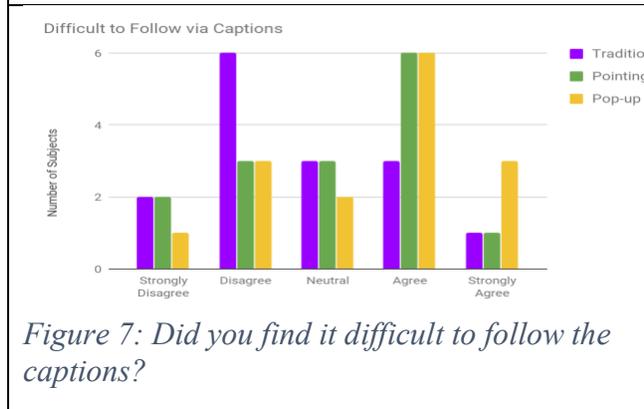

*Figure 7: Did you find it difficult to follow the captions?*

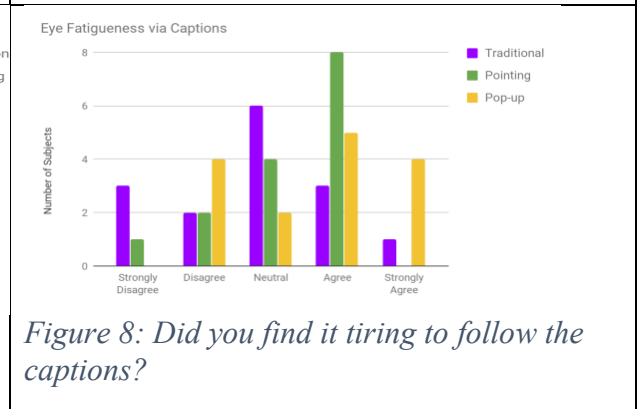

*Figure 8: Did you find it tiring to follow the captions?*

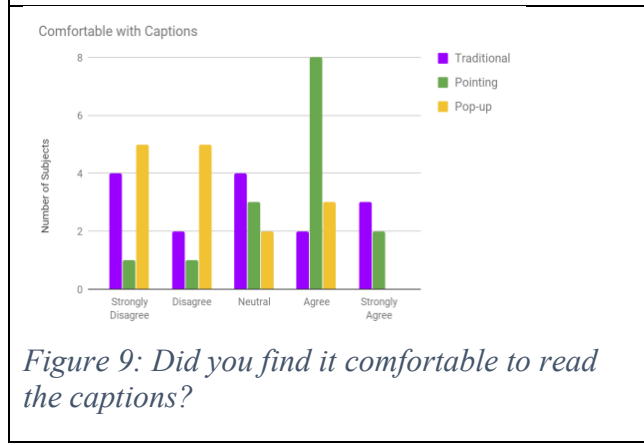

*Figure 9: Did you find it comfortable to read the captions?*

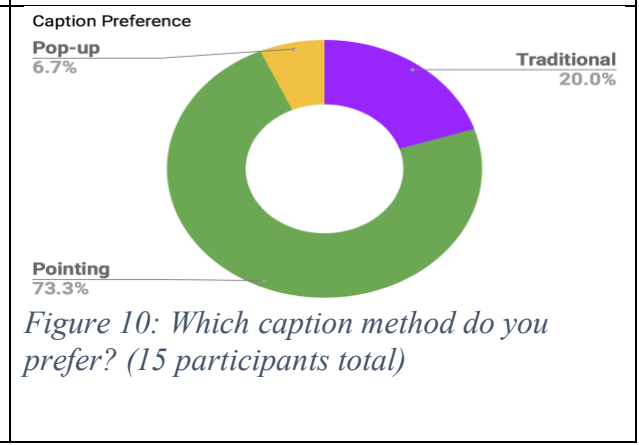

*Figure 10: Which caption method do you prefer? (15 participants total)*

**Discussion**

Based on their responses to the question about identifying the speaker, more than half of participants strongly disagreed that they can identify the speaker through traditional captioning. The responses also indicated that the pop-up method has a balance between agree and disagree, while the pointing method had most responses marked strongly agree. While participants who preferred pop-up captions could identify the speaker through pop-up captioning, they did not like it when the captions disappear and reappear in a different place. It was difficult to read the moving words.

More than two-thirds of participants agreed that they could identify the speaker through pointing caption. One of participants' positive comment about the pointing method:

*"Best one of all. I like how there was an indicator that should who was speaking. Very beneficial."*

Pop-up captions were disliked the most by the participants, according to Figure 10. Only 6% of participants preferred the pop-up caption over two other captions. Only 20% of participants preferred the traditional caption over the others. Overall, 73% of participants liked the pointing captions more than others, in part because the viewers could rely on captions being in one place, while still being able to identify the speaker through the pointer was as evidenced by the following comment by a participant:

*"It shows who is speaking and it stays in one place."*

We compared each captioning method, using a chi-square test. For readability and fatigue, no significant differences were found. For acquiring information and comfortable, only the comparison between the pop-up and pointing motions had a significant difference. For identifying speakers, all three comparisons had P-Values less than 0.05. Therefore, all three comparisons had significant differences in terms of identifying the speaker.

For ease of acquiring information in reading captions, the only significant difference was between pop-up and pointing. This is probably because text is designed to be static, so when captions (text representation of speech) moves quickly among multiple speakers, it can be difficult for viewers/readers to follow and read the captions.

| Captioning Style | Identify Speaker | Acquire information | Easy to Follow | Eye Fatigueness | Comfortable |
|---|---|---|---|---|---|
| Traditional vs. Pop-up | **0.0002** | 0.1306 | 0.1109 | 0.0753 | 0.2587 |
| Traditional vs. Pointing | **<0.0001** | 0.3840 | 0.3047 | 0.1306 | 0.1189 |
| Pop-up vs. Pointing | **0.0044** | **0.0359** | 0.2711 | 0.3343 | **0.0017** |

Table 1: This table demonstrates the different p-values when comparing to α=0.05. The red colored font means there is not a significant difference whereas the bold blue colored font means that there is a significant difference.

**Conclusion**

The pointing method is the most preferred by the participants, while the traditional method is the least preferred, as shown in Figure 10. The pointing method display is like traditional method display because the captions remain in one place. So, deaf viewers can read the caption in one place while being directed toward the speakers who talk using the pointing feature. This feature can also help hearing audience members, to identify the speaker in panel discussions.

Research indicates that a deaf individual's peripheral vision affects attention more acutely than a hearing individual's peripheral vision, as hearing people tend to focus more on the center of their field of vision [15]. This finding suggests that in multiple speaker settings where deaf viewers already redirect their gaze to various focal points, deaf viewers may benefit from peripheral cues such as a pointing arrow from captions to speaker or from captions that "pop-up" next to an active speaker.

**Future work**

Additional steps to further improve the real-time tracking display of captions could include incorporation of new captioning display methods like displaying the captions in augmented reality systems.

Other possible improvements could include the use of eye-tracking devices to reduce eye-movement fatigue when using this technology, and to expand upon the sample size for future studies.

**Acknowledgements**

This material is based upon work supported by NSF Awards #1460894 and #1757836.